# A Consensus Algorithm Based on Risk Assessment Model for Permissioned Blockchain

Xiaohui ZHANG[①]    Mingying XUE[*③]    Xianghua MIAO[①②]

[①](*Faculty of Information Engineering and Automation, Kunming University of Science and Technology, Kunming, China*)

[②](*Computer Technology Application Key Laboratory of Yunnan Province, Kunming, China*)

[③](*Faculty of Management and Economics, Kunming University of Science and Technology, Kunming, China*)

**Abstract**. Blockchain is characterized by privacy, traceability, and security features as a novel framework of distributed ledger technologies. Blockchain technology enables stakeholders to conduct trusted data sharing and exchange without a trusted centralized institution. These features make blockchain applications attractive to enhance trustworthiness in very different contexts. Due to unique design concepts and outstanding performance, blockchain has become a popular research topic in industry and academia in recent years. Every participant is anonymous in a permissionless blockchain represented by cryptocurrency applications such as Bitcoin. In this situation, some special incentive mechanisms are applied to permissionless blockchain, such as "mined" native cryptocurrency to solve the trust issues of permissionless blockchain. In many use cases, permissionless blockchain has bottlenecks in transaction throughput performance, which restricts further application in the real world. A permissioned blockchain can reach a consensus among a group of entities that do not establish an entire trust relationship. Unlike permissionless blockchains, the participants must be identified in permissioned blockchains. By relying on the traditional crash fault-tolerant consensus protocols, permissioned blockchains can achieve high transaction throughput and low latency without sacrificing security. However, how to balance the security and consensus efficiency is still the issue that needs to be solved urgently in permissioned blockchains. As the core module of blockchain technology, the consensus algorithm plays a vital role in the performance of the blockchain system. Thus, this paper proposes a new consensus algorithm for permissioned blockchain, the Risk Assessment-based Consensus protocol (RAC), combined with the decentralized design concept and the risk-node assessment mechanism to address the unbalance issues of performance in speed, scalability, and security.

## 1. INTRODUCTION

Bitcoin is a type of digital money that has taken the world by storm. The concept of bitcoin was first proposed by a scholar in 2008, under the pseudonym of Satoshi Nakamoto, and then the Bitcoin system was released on the Internet [1]. There are no centralization management servers and third-party credit endorsement organizations. Bitcoin has been operating stably for more than ten years. These characteristics demonstrate the potential advantage of the technology behind the Bitcoin system, blockchain. Blockchain is a decentralized distributed ledger that generates and stores data in blocks and constructs a chain structure in chronological order. The security and immutability of blockchain are grounded in cryptography, smart contract, P2P network, and consensus protocol [2]. Blockchain has brought hope to solve the issues such as privacy, security, and trustworthiness in distributed ledger technologies [3]. In recent years, more and more researchers have applied blockchain to other scenarios, such as supply chain [4], information security [5], data security [6], Internet of things [7], etc. With the development of blockchain technology, a new business model is emerging, based on blockchain to help a group of entities to get rid of the dependence on centralization certification organization [8]. It is foreseeable that blockchain will gradually become an indispensable part of the future Internet.

There is no central organization to undertake the data verification work, every stakeholder must record the same correct data because a peer-to-peer architecture was used in the decentralized distributed system like blockchain. However, it is difficult to maintain the same content and sequence of transactions in all the participants due to the different status of the participants and the network environment in which they are located [9]. In addition, some participants may be attacked as byzantine nodes to obstruct transmission. Therefore, blockchain technology also brings new problems to the system, the solution to these problems relies heavily on consensus algorithm because consensus algorithm plays a vital role in the performance of the blockchain system [10].

For example, permissionless blockchain such as Bitcoin employs the methods that "mined" the cryptocurrency using their computing powers to mitigate the absence of trust. PoW (Proof-of-Work) [1], PoS (Proof of Stake) [11], and DpoS (Delegated Proof of Stake) [12] are classified as permissionless blockchain consensus protocols. However, PoW has limitations in computing power consumption and small throughput. In addition, the PoW consensus may suffer the tailored attack behavior such as 51% attacks [13]. Although the PoS and DPoS solve the waste of resources in PoW, there are still problems such as low efficiency [14]. Permissioned blockchain is more suitable for high real-time applications due to its high transaction throughput performance and low transaction confirmation latency [15]. Permissioned blockchain can use classic consensus algorithms, such as Raft [16] and PBFT [18], to reach the consensus among entities because all participants must be identified. PBFT has not been widely used in real-world projects due to high consensus cost and poor scalability performance in many use cases. Raft can only be used in non-byzantine environments that only honest nodes in the network [19]. The above examples all show that the many consensus algorithms unable to meet Quality of Service (QoS) demands of specific scenarios is an essential reason that hinder the broad application.

Therefore, in order to meet the requirements of high scalability and security as much as possible under the premise of decentralization performance, a new consensus algorithm for Permissioned blockchain put forward in this article appropriable, which is called RAC (Risk Assessment-based Consensus protocol). The main contributions of our work are as follows.

(1) There is no centralized endorsement organization in the blockchain system. The participant with strong computing power or high-stake rights is often used to be the accountant node that packs transactions into a block and sends the block to other participants, which weakens the decentralized characteristics of the blockchain system. We have designed a new decentralized consensus model to avoid monopolistic behavior caused by excessive concentration of authority. In our model, the achievement of consensus relies on cooperation between all roles. At the same time, different roles can achieve conversion under certain conditions.

(2) It will have an immeasurable impact on the blockchain system when the accountant node is maliciously controlled. We have designed an efficient accountant node election strategy to ensure that only honest participants can act as the accountant to generate new blocks, combined with the risk-node assessment mechanism to identify the malicious nodes that may exist in the network.

(3) The distributed connectivity of the blockchain exposes the systems to byzantine attacks. The compromised participants will further decrease the trust level among cooperative organizations by generating false data to obstruct consensus. We have designed an efficient block addition and transaction confirmation strategy to prevent possible Byzantine behavior and collusion attacks in the network. The process achieves a credible consensus by combining the reward and punishment mechanism and addressing the unbalance issues of performance in speed, scalability, and security in the consensus algorithm proposed before.

The remainder of this article is organized as follows. Chapter 2 reviews the research related to the blockchain consensus algorithm. Chapter 3 introduces the basic definition and system model of the RAC algorithm. Chapter 4 describes the detailed implementation process of the RAC algorithm. Chapter 5 demonstrates the performance of the RAC algorithm through theoretical analysis and experiments. Chapter 6 summarizes the research work and provides an outlook for future work.

## 2. RELATED WORK

There have been several works about blockchain and consensus algorithms in recent years, and our work on the consensus model is motivated by some studies proposed before. In this section, we present a brief literature review from three aspects.

### 2.1 Overview of Blockchain

A blockchain is a tamper-evident ledger supported by a consensus algorithm. Peer to Peer (P2P) networks, cryptographic hash, digital signatures, and smart contracts are core blockchain technologies. Every node in the blockchain maintains a copy of the ledger verified by a consensus protocol, as shown in Fig.1. The ledger exists in blocks, each of which is linked to the previous block by a hash[20]. Since each node in a P2P network has equal status, there is no need for a central server to exchange information in the blockchain system[21]

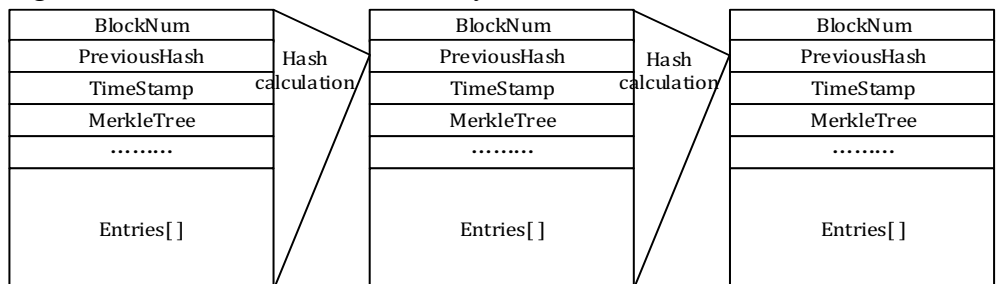

**Fig.1. The structure of block in consensus algorithm**

### 2.1.1 Blockchain transaction process

As shown in Fig.2., the transaction process of any blockchain system can be classified into three stages: Accountant selection, Block addition, and Transaction confirmation [14].

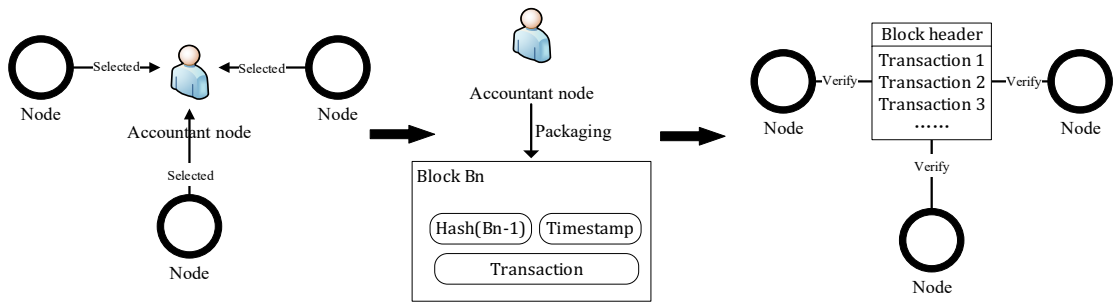

Fig.2. the transaction process of blockchain

Accountant selection is the first stage of the transaction process. Accountant nodes are responsible for collecting and generating blocks from the client, then packaging transactions into blocks and sending blocks to other nodes in the blockchain network. Accountant nodes can be generated by competing (e.g., PoW, PoS), polling (e.g., PBFT), or voting (e.g., Raft) approaches.

Block addition is the second stage of the transaction process. Since each node maintains a copied ledger in local, a node adds the block into the local ledger after the block verification is passed. In a few consensus algorithms, such as PBFT, block addition requires a majority vote of the nodes. Therefore, these consensus algorithms generally suffer from a lack of scalability and high consensus cost.

Transaction confirmation is the third stage of the transaction process. The purpose of transaction confirmation is to confirm the transaction's validity based on the blockchain held by each node in the blockchain network. However, most consensus algorithms do not require real-time voting by the nodes during the block addition phase. Hence, the confirmation efficiency relies on the consensus algorithm's characteristics used in the blockchain system. For example, confirming a transaction in the Raft algorithm requires that more than 50% of the nodes complete the block addition.

### 2.1.2 Types of blockchains

Blockchain can be divided into permissionless and permissioned [17]. Bitcoin, Ethereum, and other cryptocurrencies issued to the public have been divided into permissionless blockchains. Any anonymous individual can freely join and maintain the blockchain network. There is no trust until the blockchain state reaches an immutable block depth. The measures of "mining" to provide financial incentives were be used to compensate for the trust issues and offset Byzantine fault-tolerance costs. So, there are weak concurrent transaction capability, high transaction fees, and long confirmation time in the permissionless blockchain. In addition, participants in the permissionless blockchain are impossible to be identified.

However, some requirements must be considered in many cases, such as transaction throughput performance, transaction confirmation latency, the confidentiality of transactions, and participant identity[22]. These characteristics are the performance requirement of use cases and regulations that must be followed. Know-Your-Customer (KYC) is one of the essential principles in an actual financial transaction, which cannot be achieved in a permissionless blockchain.

Permissioned blockchains, on the other hand, provide a way to achieve the interactions between a set of entities that do not establish an entire trust relationship.

Since the participants' identities are known, permissioned blockchains can use a more efficient consensus algorithm to drive smart contracts. Such as crash fault tolerant (CFT) or byzantine fault tolerant (BFT) protocols. As a result, permissioned blockchain has lower transaction costs and higher efficiency than permissionless blockchain, which uses an "incentive mechanism" to reach a consensus.

Table 1 compares permissionless blockchains and permissioned blockchains in terms of various parameters. To sum up, transactions are executed on every node in the permissionless blockchain network, this drawback can be problematic for many use cases, such as supply-chain, securities industry, because none of partners would want their competitors to know confidentiality information in these businesses. On the contrary, permissioned blockchain system have adopted a variety of approaches to address the lack of privacy and further improve the efficiency, these characteristics make permissioned blockchain acceptable in many enterprises use cases.

Table 1. Comparisons between permissionless blockchain and permissioned blockchain

| Sl. No. | Criteria | Permissionless | Permissioned |
|---|---|---|---|
| 1 | Decentralization | High | Low |
| 2 | Identity | Anonymous | Identifiable |
| 3 | Award | Mostly yes | Mostly no |
| 4 | User scale | Large | Small |
| 5 | Classic project | Bitcoin | Hyperledger |

### 2.1.3 The challenge in blockchain based applications

Permissionless blockchains are more suitable for crypto-currency applications in open and anonymous environments. The number of users in such a case is enormous. On the other side of the coin, permissioned blockchains are more suitable for cooperation between a certain scale of organizations without a centralized authority party. All stakeholders of permissioned blockchain must be identified and identifiable.

Although blockchain has been applied in many aspects, including in the area of IoT, smart city, supply chain, vehicular ad-hoc networks, and so on, it is still a technology that is under development, which means there are some obstacles exist in blockchain-based applications need to be and improved[20]. From what has been analyzed in some previous work, the contradiction between Decentralization, Security, and Scalability is the main reason that restricts the development of blockchain-based applications.

The three properties, consistency, availability, and partition tolerance can only satisfy most two simultaneously in a distributed system, which was named the "CAP theorem" and proved in 2002 by Seth Gilbert and Nancy Lynch[23]. CAP provides a guiding principle for the design of consensus algorithms. From this, researchers no longer pursue that can satisfy all three properties simultaneously. Based on the CAP theorem, Vitalik Buterin proposed the DSS conjecture in the Ethereum system: the blockchain system cannot be enhanced in terms of Decentralization, Security, and Scalability at the same time[24]. From the DSS conjecture, we can conclude that the transaction throughput of permissioned blockchains where total decentralization is not provided is much higher than permissionless blockchains. For example, a decentralized bitcoin system is much smaller than Hyperledger Fabric with partial decentralization in throughput performance.

Decentralization means that the accountant node is fairly generated throughout all nodes in the blockchain network rather than being centralized by a small number of users. The degree of decentralization and the control of the blockchain network are closely related. It is essential to ensure the fair participation of users.

Security means the main security issues and corresponding protection mechanisms in blockchain systems, including cryptographic security, network security, data security, consensus algorithm security, etc. These include Collusion attack, Double Spending attack, Sybil attack, and Targeted Attack.

Scalability refers to the ability of the blockchain system to process the transaction. Three main aspects influence the performance of scalability. First, network latency of the distributed system, the network latency of a distributed system is more limited than that of a traditional distributed system due to an uncontrollable network environment. Second, the consistency of all users in blockchain requires necessary confirmation computation to maintain the consistency among nodes, resulting in additional computational cost. Third, constraints in computational performance of blockchain nodes, especially in IoT applications, many nodes cannot process some consensuses that needs a large amount of energy and computational consumption, such as PoW.

Although the DSS conjecture is only a way to analyze the blockchain performance of Ethereum, it is not theoretically rigorous. Still, it provides an entry point for researchers to improve the performance of the consensus algorithm. From what has been analyzed above, developing suitable consensus algorithms to balance the performance between Decentralization, Security, and Scalability is the key to solving the challenge in blockchain-based applications[25].

**2.2 Proof-Based Consensus Algorithm**

PoW [1] algorithm is used in permissionless blockchain systems. Participants compete for accounting rights by solving a complex but easily verifiable mathematical problem with their computing power. The first node to solve such a problem is rewarded with a certain amount of cryptocurrency. This process is described as "mining," and the feature of open access makes PoW more scalable and decentralized. However, there is the problem of computing power consumption in PoW. It is unfair that the competition for accounting rights is closely related to the computing power of nodes because nodes with low CPU capacity is difficult to obtain accounting rights of PoW. In addition, a transaction in a block needs to wait for the confirmation of six additional subsequent blocks before consensus can be achieved to prevent double-spending attacks in the consensus process. Studies have shown that the PoW algorithm can only reach consensus for seven transactions per second and may suffer the tailored attack behavior such as 51% attacks. The low consensus efficiency limits the application of PoW in blockchain systems other than Bitcoin [26].

Researchers have proposed the PoS consensus to solve the problem of computing power consumption in the PoW algorithm, which has been applied in Ethereum [11]. In the PoS algorithm, each node is given a new metric called coinage. PoS algorithm assumes that rational high-asset nodes will not disrupt the consensus process because the potential loss of assets will outweigh the gain from their malicious deeds. The PoS algorithm achieves higher TPS performance compared to the PoW algorithm, but there are also shortcomings. The risk of monopoly is introduced in PoS. Low-asset nodes cannot compete fairly with high-asset nodes for accounting rights, and it can lead to a tendency to centralize the system. Researchers have proposed the DPoS algorithm to mitigate this monopoly risk in PoS, which is currently used in projects such as EOSIO and Cosmos [12]. Each node can vote for a representative based on their stake, and the accounting rights will be given to the node that receives the highest number of votes. Although the PoS and DPoS solve the waste of resources in PoW, there are still problems such as low efficiency.

Proof-based consensus algorithm is designed to crypto-currency system such as Bitcoin and Ethereum. Although the efficiency of proof-based consensus algorithm is low, it is worth mentioning that these algorithms make some attack behavior for blockchain system impractical. We provided a summary comparison in Table 2 about its pros and cons.

Table 2 Comparisons between classic proof-based consensus algorithm

| Sl. No. | Consensus algorithms | Classic application | Decentralization level | Energy efficiency | Proof Based on |
|---|---|---|---|---|---|
| 1 | PoW | Bitcoin | Decentralized | No | Work |
| 2 | PoS | Ethereum | Semi-centralized | Yes | Stake |
| 3 | DPoS | Ethereum | Semi-centralized | Yes | Vote |

**2.3 Voting-Based Consensus Algorithm**

Permissioned blockchain has higher requirements for transaction throughput than permissionless blockchain, and it is more suitable to adopt the voting-based consensus mechanism. In this section, we will introduce some voting-based consensus mechanism and their improvement algorithms. The voting-based consensus mechanism can be divided into Byzantine Fault Tolerance algorithms that can tolerate Byzantine Failure and Crash Fault Tolerance algorithms that can only handle Fail-stop Failure. Byzantine General Problem proposed by Lamport et al. in 1982 is how the participants can complete consensus in the presence of malicious node interference in the network [27].

The differences between the two consensus algorithms are shown in Table 3. It should be noted that all the data of TPS in the Table are the result under ideal experimental conditions.

Table 3 Comparisons between two types of consensus algorithm in permissioned blockchain

| Sl. No. | Criteria | CFT algorithm | BFT algorithm |
|---|---|---|---|
| 1 | Crash fault tolerance | 50% | 33% |
| 2 | Byzantine fault tolerance | N/A | 50% |
| 3 | Throughput (TPS) | >10k | >1k |
| 4 | Classic algorithms | RAFT | PBFT |

**2.3.1 Byzantine Fault Tolerance algorithms**

As a classical BFT algorithm based on the state machine replication mechanism, the PBFT algorithm was proposed by M. Castro and B. Liskov et al [27]. PBFT algorithm can reach consensus with the number of malicious nodes accounting for less than 1/3. However, the PBFT algorithm requires mutual communication and confirmation of every two nodes in each consensus, and the algorithm's complexity is $O(n^2)$. Thus, the application scenario of PBFT is minimal due to its high consensus cost and poor scalability performance. PBFT has not been used in real-world engineering applications. For example, the famous open-source project of permissioned blockchain, Hyperledger Fabric, has only used the PBFT algorithm in the early v0.6 version.

Lei et al. proposed a reputation-based Byzantine fault-tolerant algorithm, RBFT, in 2018[28]. RBFT evaluates the behavior of each node in the consensus process through a reputation model. If a malicious node is detected, its reputation value and voting weight will be reduced. In addition, the view change phase is improved by adding an incentive mechanism so that the nodes with higher reputation value have a higher chance to become the accountant node. The experimental results show that the RBFT algorithm can quickly and effectively identify malicious nodes through the reputation model. Compared with PBFT, the RBFT algorithm can eventually improve the transmission throughput by 15% and reduce the latency metric by 10%.

Rong et al. proposed an improved BFT algorithm, ERBFT, in 2019[29]. ERBFT algorithm uses the backup nodes in the network to identify whether the accountant node is a malicious node through a new request ordering mechanism (Order-Match). Suppose the accountant node does not pass the verification of the Order-Match mechanism. In that case, the backup node will trigger the suspicious protocol (suspect protocol) to confirm further whether the accountant node is under malicious attack. The experimental results show that the ERBFT algorithm performs better than the PBFT algorithm in throughput and scalability and can improve the transaction throughput by 30%.

Tendermint is a blockchain project of the Cosmos network, known for simplicity, high performance, and fork accountability. As the consensus protocol of the Tendermint, Tendermint BFT is a simplified version of the PBFT algorithm, the relationship between Tendermint BFT and PBFT is similar to the relationship between Raft and Paxos. Tendermint BFT can be used in a Byzantine environment. Although Tendermint BFT can tolerate the failure of only 1/3 of the nodes in the system, it can tolerate any fault, including hacking and malicious attacks [30].

The core idea of the above study can be summarized as optimizing the traditional Byzantine fault-tolerant algorithm by improving the transaction throughput and scaling performance. Although the current research results have improved the performance and scalability of the traditional Byzantine fault-tolerant algorithm, the transaction efficiency is still far from that of the Crash Fault Tolerance algorithm.

**2.3.2 Crash Fault Tolerance algorithms**

Raft algorithm [16] is an improvement of the Paxos algorithm, Raft algorithm completes the consensus through the leader election phase and logs replication phase. The nodes of the Raft cluster do not need to confirm each other for the transmitted data. The transmission throughput of the Raft algorithm in the experimental environment is more than five times that of the PBFT algorithm. The Raft algorithm has been used in many practical projects such as ETCD and BRAFT due to their easy-to-understand and excellent performance. However, the Raft algorithm can only be used in a non-Byzantine network environment without malicious nodes, which cannot restrict the behavior of malicious nodes, so it will bring incalculable impact to the whole consensus system once the leader node is maliciously controlled [19].

Chen et al. proposed a CRaft consensus algorithm based on the node trust mechanism based on the Raft algorithm in 2018[31], which enables the algorithm used in Byzantine network environment. CRaft is divided into trust evaluation phase and consensus phase. It establishes trust evaluation criteria by OC-SVM algorithm. The prediction accuracy of Craft for Byzantine nodes is up to 100% (there is still a 17.89% false-positive rate in CRaft). Compared with the PBFT algorithm, the throughput of CRaft can still be maintained at good performance when the number of nodes is expanded to 60. So, it is more suitable for a permissioned blockchain environment than PBFT.

Wang et al. developed the Beh-Raft algorithm in 2021, which divides blockchain nodes into groups as parallel shadings to improve scalability at the cost of increased communication and storage per node [32]. Beh-Raft introduces the Proof of Behavior algorithm (PoB) for incentivizing honest behavior, which ensures that the probability of an honest node being selected as an accountant node is greatly increased by combining the PoB with the Raft algorithm. The Beh-Raft algorithm is Byzantine fault-tolerant while maintaining better scalability.

The hhraft algorithm was proposed by wang et al. in 2021[33]. In response to the problem that the RAFT algorithm cannot use in a real-world network, hhraft introduces a new monitor mechanism to optimize the Raft consensus process, which is used to monitor the network for "Sybil nodes" that fake their identities maliciously and accountant nodes that tamper with the original data. The hhraft has been experimentally proven to outperform the Raft algorithm in terms of transaction throughput, consensus latency, and resistance to Byzantine failures. It is suitable for real network environments with high real-time and high adversarial performance.

Although the current research results can determine the malicious nodes in the network, there are still some shortcomings. Most of the current improved algorithms determine malicious nodes in a "static" method, which means they need to design malicious node models or reputation value determination criteria in advance and cannot be adjusted according to network environment changes dynamically. However, the imbalance between the behavior of normal nodes and malicious nodes makes it difficult to construct an accurate classifier in a Byzantine network environment.

## 3. THE SYSTEM MODEL
### 3.1 Problem Description

In order to realize some business goals without centralized institutions and meet the demands of traceability and verifiability, such as data sharing or commercial paper exchange, a permissioned blockchain is usually established by several companies or organizations with a full trust relationship not established between them. In this blockchain network, everyone wants to have administrative privileges because everyone fears losing rights and profits in decentralized networks. It is an acceptable solution for all participants to monitor each other to prevent the organization that masters the power to dominate the system from gaining illegal benefits. In such a case, the participants do not subjectively destroy to cause byzantine behavior to the permission blockchain unless they are maliciously controlled because everyone wants to benefit from the blockchain system. In addition, the participants are abstracted as nodes. Nodes are composed of computational devices such as sensors, computers, and servers in the real environment. The lowest possible consumption of computational resources is what every stakeholder of the permission blockchain expects to achieve due to computational resources, and economic costs are closely linked. Therefore, it is essential to find a suitable consensus algorithm to meet the demand of such a consortium blockchain.

From what has been analyzed in section 2.1, the evaluation system of the consensus algorithm can be established based on aspects of decentralization, security, and scalability. More detailed, each evaluation criteria can be measured by several sub-indicators[25], so the main parameters of the evaluation consensus algorithm system can be summarized in Table 4.

Table 4. The evaluation system of consensus algorithm

| Sl. No. | Criteria | Sub-indicators | Description |
|---|---|---|---|
| 1 | Decentralization | Number of consensus nodes | The number of nodes in the blockchain network that can become the accountant node. |
| 2 | Decentralization | Accountant selection method | The method that accountant node is generated in the blockchain network, competition, election, or polling. |
| 3 | Decentralization | Consensus nodes weight | Whether the consensus nodes have equal probability of becoming the accountant node in each consensus process. |
| 4 | Security | Byzantine fault tolerance | The maximum percentage of Byzantine malicious nodes in the whole blockchain network that can be accepted. |
| 5 | Security | Byzantine node controllability | The ability of consensus algorithm to exclude malicious nodes from the consensus process |
| 6 | Security | Attack behavior costs | The cost of an attack behavior by a malicious attacker in the blockchain network |
| 7 | Scalability | Resource consumption | Resources to be consumed in the consensus process include computation, storage, and network communication, which can be measured by the communication complexity of consensus. |

Number of consensus nodes: In some consensus algorithms, not all nodes can participate in the election of accountant nodes because some consensus nodes need to undertake other tasks such as supervising the consensus process's realization. On the contrary, all nodes can be selected as accountant nodes due to no additional roles in the consensus algorithm. Number of consensus nodes is the Qualitative and Beneficial Indicator, we set $\{0,1\}$ to indicate the $\{Part, All\}$ respectively.

Accountant selection method: In general, voting and polling are two of the most widely used in permissioned blockchain. Compared to polling, the voting is somewhat less decentralized because nodes may not have an equal probability of getting a vote under different network environments. Accountant selection method is the Qualitative and Beneficial Indicator, we set $\{0, 0.5, 1\}$ to indicate the $\{Competing, Voting, Polling\}$ respectively.

Consensus nodes weight: The probability of a consensus node that is honest becoming an accountant node depends on the presence of constraints. For example, the credit evaluation standard that includes the node performance is set in CRAFT. Only the node that meets this credit evaluation standard can be selected. However, some consensus algorithm does not have constraints, which means that honest nodes have the same probability of being the accountant node. Consensus nodes weight is the Qualitative and Beneficial Indicator, we set $\{0, 1\}$ to indicate the $\{Not\ equal\ weighting\ of\ consensus\ nodes, Equal\ weighting\ of\ consensus\ nodes\}$ respectively.

Byzantine fault tolerance: The authors summarize this metric in their respective papers. Byzantine fault tolerance is the Qualitative and Beneficial Indicator. We assign a value to Byzantine fault tolerance with reference to the following criteria.

Table 5. Scoring basis for Byzantine fault tolerance

| Value | Range of byzantine fault tolerance | Description |
|---|---|---|
| 0 | 0% | Consensus algorithm that completes failure to recognize Byzantine behavior in the network. |
| 0.3 | 1% - 16% | Consensus algorithm that has ability to tolerate a small number of Byzantine nodes in the network. |
| 0.5 | 17% - 33% | Consensus algorithm that has ability to tolerate several numbers of Byzantine nodes in the network. |
| 0.7 | 34% - 51% | Consensus algorithm that has ability to tolerate no more than half of Byzantine nodes in the network. |
| 1 | >51% | Consensus algorithm that can withstand 51% of attacks. |

Byzantine node controllability: In some consensus algorithms, identified malicious nodes can continue to compete for the accountant node. Reducing the impact of malicious nodes is one of the essential metrics for measuring security. In RAC, reducing the weight of malicious nodes can also prevent them from becoming accountant nodes because they cannot get more than half of the votes under the Risk-Node Assessment Mechanism. Byzantine node controllability is the Qualitative and Beneficial Indicator, we set $\{0, 1\}$ to indicate the
$\{No\ controllability\ over\ Byzantine\ nodes, Having\ control\ over\ Byzantine\ nodes\}$ respectively.

Attack behavior costs: This metric indicates the cost consumed by attackers needed to malicious control the whole permissioned blockchain. Attack behavior costs include the following two areas: malicious control of the accountant node or enabling permissioned blockchain to reach consensus under the attackers' intentions. Attack behavior costs are the Qualitative and Beneficial Indicator. We assign a value to Attack behavior costs regarding the following criteria.

Table 6. Scoring basis for Attack behavior costs

| Value | Attack behavior costs | Description |
|---|---|---|
| 0 | None | An attacker can achieve malicious control of the blockchain system at no cost. Such as solo consensus of Hyperledger Fabric. |
| 0.3 | Low | The attacker only needs to spend a small amount of cost to control critical parts such as accountant or node with evaluation function to achieve malicious control of the blockchain system. Such as Raft. |
| 0.6 | Middle | The attacker only needs to spend many costs to achieve malicious control of the blockchain system because a certain malicious attack detection and prevention mechanism is set in the consensus algorithm. |
| 1 | High | It is almost impossible for an attacker to achieve complete control of the blockchain system due to the very strict confirmation mechanism, unless maliciously controlling more than half of the nodes, such as PoW and PBFT. |

Resource consumption: The communication complexity of consensus can measure this metric. The authors summarize the communication complexity of proposed consensus algorithms in their respective papers. Resource consumption is the Qualitative and Cost Indicator. We assign a value to Resource consumption regarding the following criteria.

Table 7. Scoring basis for Resource consumption

| Value | Resource consumption | Description |
|---|---|---|
| 0 | $> O(n^2)$ | Extremely resource-intensive consensus algorithms. |
| 0.3 | $O(n^2)$ | Consensus algorithms that require complex confirmation mechanisms, such as PBFT. |
| 0.5 | $O(nlogn)$ | Consensus algorithms that require some additional confirmation mechanism, such as Beh-Raft. |
| 0.7 | $O(n)$ | Consensus algorithm with efficient confirmation mechanism, such as Raft. |
| 1 | $O(1)$ | Ideal-state consensus algorithm, typically used in experimental settings, such as Solo consensus of Hyperledger Fabric. |

We can describe the problem that development of appropriate consensus algorithm based on some indicators as a mathematical model. Let eigenvector $S = \{S_1, S_2, \ldots, S_n\}$ indicates to the consensus algorithms that need to be compared. Assume that there $m$ ($m = 7$ in the case analyzed before) indicators in evaluation system, then $a_{ij}$ ($i = 1,2,\ldots,n; j = 1,2,\ldots,m$) is denoted as the value of corresponding indicator observation of decentralization, security and scalability. The evaluation indicator is preprocessed by normalization and dimensionless method, and the evaluation matrix $\boldsymbol{B} = (b_{ij})_{n \times m}$ is constructed.

Positive ideal solution $C^+ = [c_1^+, c_2^+, \ldots, c_m^+]$ denotes an ideal consensus algorithm that not existed in real-world (each indicator of $C^+$ is the best value among all the consensus algorithm). Set the value of the $j$th indicator in $C^+$ is $c_j^+$,

$$c_j^+ = max\ (b_{ij}),\ 1 \leq i \leq n\ and\ 1 \leq j \leq m \tag{1}$$

On the contrary, Negative ideal solution $C^- = [c_1^-, c_2^-, \ldots, c_m^-]$ denotes another ideal consensus algorithm (each indicator of $C^-$ is the worst value among all the consensus algorithm). Set the value of the $j$th indicator in $C^-$ is $c_j^-$,

$$c_j^- = min\ (b_{ij}),\ 1 \leq i \leq n\ and\ 1 \leq j \leq m \tag{2}$$

Our objective is to find the consensus algorithm that has maximum proximity between ideal solution. Therefore, the objective problem is defined as equation (3).

$$max\ \{f_i = s_i^- / (s_i^- + s_i^+)\},\ 1 \leq i \leq n \tag{3}$$

where $s_i^+$ and $s_i^-$ are calculated as follows.

$$s_i^+ = \sqrt{\sum_{j=1}^{m}(b_{ij} - c_j^+)^2},\ 1 \leq i \leq n \tag{4}$$

$$s_i^- = \sqrt{\sum_{j=1}^{m}(b_{ij} - c_j^-)^2},\ 1 \leq i \leq n \tag{5}$$

### 3.2 Basic Definitions

The participants of permissioned blockchain are abstracted as nodes, either from the same organization or from different organizations, and will carry a certain amount of assets or stake interests. First, we need to introduce some basic definitions to help the reader better understand the proposed consensus algorithm.

(1) Byzantine node and honest node: Byzantine nodes are defined as nodes that malicious attackers may compromise to hinder consensus in participants. Byzantine nodes are also referred to as risk or malicious nodes in this paper. On the contrary, the honest node indicates the node making the right decision on the consensus. The definition of a right decision is determined by the behavior of nodes under different states, The introduction of the states and behaviors will be given in section 3.3 and 4.3.

(2) RNL: The node in the Risk node list (RNL) means that the node is not trusted anymore because the Byzantine General problem or other malicious behavior has occurred in these nodes. RNL is continuously updated during the consensus process according to the behavior of every node in the network.

(3) Term: RAC uses the term as the logical timestamp to ensure that the nodes have correct timestamps under their different physical environments. The term is numbered using consecutive integers, and the term is updated to a larger term number when a new accountant is generated. The description of the term as shown in Fig.3.

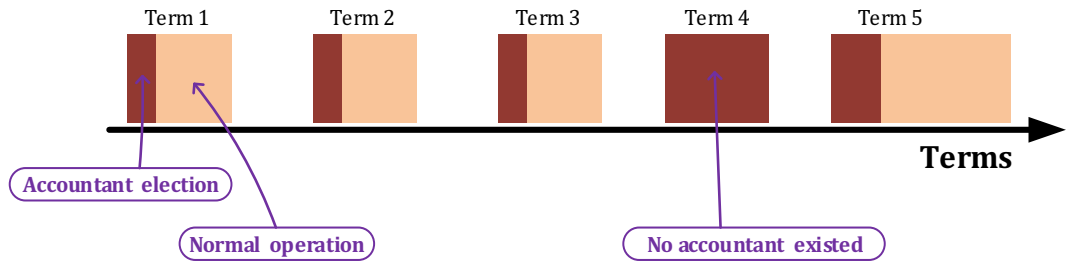

Fig 3. After a successful election, an accountant manages the cluster until the end of the term

(4) Basic operations: This definition is mainly applied in the description of the RAC process in section 4.1, which is divided into the sub-definition as follows.
- $P_i = \{P_{i1}, P_{i2}, P_{i3}, \ldots, P_{in}\}$. The set of nodes in the participating organization $i$, where $n$ denotes the total number of nodes in organization $i$.
- $E = \{E_1, E_2, E_3, \ldots, E_n\}$. The set of evaluator nodes in the permissioned blockchain network, i.e., the group of evaluator nodes.
- $O = \{O_1, O_2, O_3, \ldots, O_n\}$. The set of nodes that participating of the RAC consensus in the permissioned blockchain network, that is, the consensus node group
- $Send(Message, A, B)$. Send the message quest $Message$ from node $A$ to node $B$.
- $Pick(O, Accountant)$. An accountant node $Accountant$ is elected from the consensus node group $O$.
- $Verify(RPC, A, B)$. The interaction between node $A$ and node $B$ through an interaction function $RPC$ to complete the verification of a block, where node $A$ is the client of $RPC$ and node $B$ is the server of $RPC$.
- $Package(Message, new\_Block, Accountant)$. The accountant node $Accountant$ packages the message from client of permissioned blockchain $Message$ into a new block $new\_Block$ that can be broadcast and stored in the blockchain.
- $Re\_Blockchain(Accountant, RPC, new\_Block)$. Block update function. The accountant node $Accountant$ broadcasts the new block $new\_Block$ to all participating organizations in the network through the interactive function $RPC$.

## 3.3 The States in Consensus Algorithm

Permissioned blockchain is established by a certain number of companies or organizations. It is an acceptable solution that all participators monitor each other to prevent the organization from gaining illegal benefits. When joining the network, each node in the permissioned blockchain is verified and registered through the membership service provider and is assigned a unique ID and key pair to indicate its identity. All nodes may experience four different roles in RAC: follower, candidate, accountant, and evaluator.

(1) Follower: All nodes are labeled as followers after joining the permissioned blockchain. They can only passively receive the message about blocks that need to be added to the blockchain. When receiving requests from clients, followers need to forward these requests to the accountant for packaging because they don't have the authority to package blocks.

(2) Candidate: The candidate is the intermediate state between the follower and accountant and will not exist in the network long. Any follower node can become a candidate node when it finds that the accountant node cannot serve properly. The candidate node can initiate votes to other follower nodes and need to get the voting support of enough follower nodes before it can be converted into an accountant node.

(3) Accountant: Accountant nodes are elected from candidate nodes. Packing client requests into a block and sending them to the evaluator nodes for validation is the responsibility of the accountant node.

(4) Evaluator: Evaluator node has two tasks in the RAC. First, based on the performance of follower nodes, the evaluator node can determine whether a node to be maliciously controlled through the risk-node assessment mechanism. Second, monitor the performance of accountant nodes, and terminate the term of accountant node with abnormal behavior in time. They are high-asset nodes of each participant in the permissioned blockchain, which can primarily reduce the possibility of them actively causing damage to the blockchain system.

The state of a node is temporary, and no one can work in a specific role permanently. Different roles can be converted under certain conditions, and the conversion relationship is shown in Fig.4.

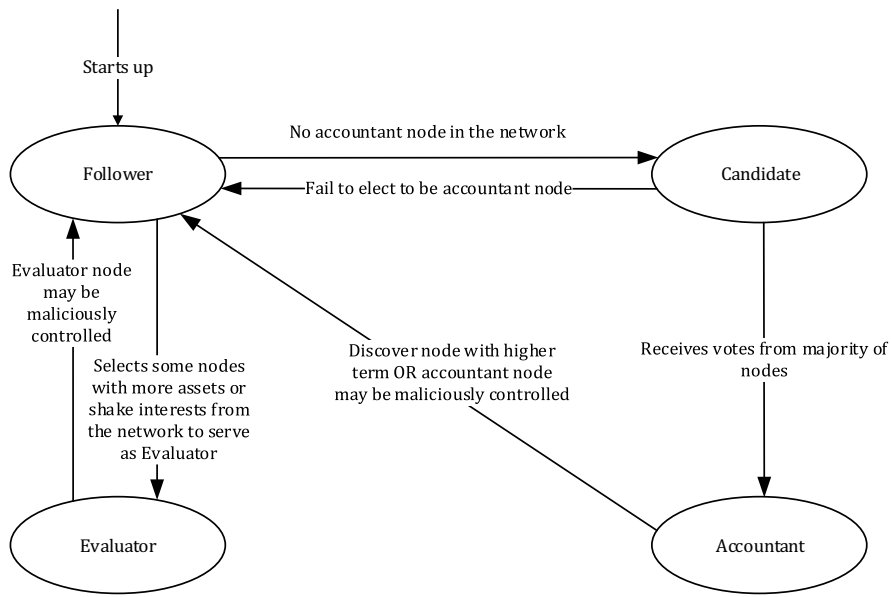

Fig 4. The conversion relationship in RAC

## 4. ALGORITHM DESIGN

## 4.1 The Description of Details in RAC Algorithm

The RAC is divided into six phases: Generation of Evaluator, Update of the RNL, Generation of Candidate and Election of Accountant, New Blocks addition, Judgment of New Blocks, New Blocks confirmation. Nodes use Remote Procedure Call (RPC) to communicate in the network. The consensus process of RAC is shown in Fig.5.

Each organization selects a certain number of nodes with more assets or computational power as evaluator nodes to represent the interests of the organization, and together they form the evaluator node group. To prevent possible collusion attacks in the subsequent consensus process, the evaluator node group needs to be composed from different organizations. $E = \{E_1, E_2, E_3, \dots, E_n\}$. Initialize the evaluator nodes of each participating organization into evaluator node groups.

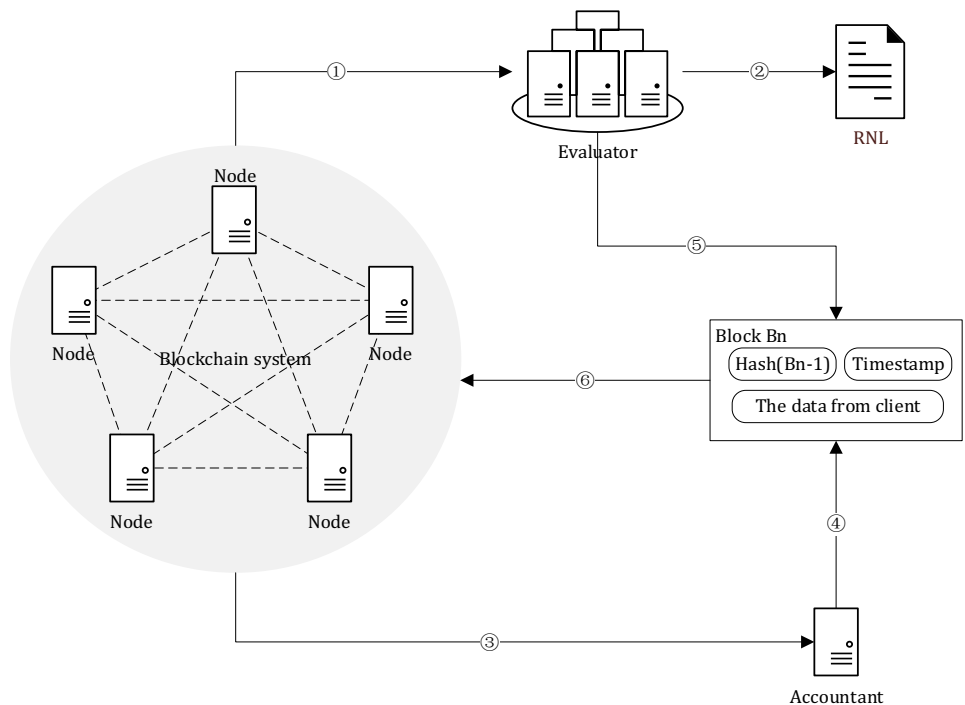

**Fig.5. The consensus process of RAC**

① $Send(SystemCall, O, E)$

All nodes send the sequence of system calls in the last term to the evaluator after the Evaluator Group is created to decide whether there are Byzantine nodes in the consensus node group that are not suitable for having accounting rights.

② $Send(RiskComputeRPC, O, E)$

The evaluator node calculates the potential risk nodes in the network through the Risk-Node Assessment Mechanism and generates and updates a list of risk nodes (RNL) to broadcast to all nodes in the network. RNL updates are implemented through RiskCompute RPC between the node and the evaluator. The description of RiskCompute RPC is shown in Table 8.

**Table 8. The description of RiskCompute RPC in the RAC algorithm**

| | RiskCompute RPC |
|---|---|
| **Parameter** | **Description** |
| Term | The current term of the node. |
| NodeID | The id of the node. |
| SystemCall | The systemcall of the node in last term. |
| **Return** | **Description** |
| Term | The current term of the node. |
| RNL | The risk node list in current term. |

③ $Pick(O, Accountant)$

A follower node that is not selected as Evaluator does not receive a heartbeat message from the accountant node in a period (randomly selected in the interval of 150-300 milliseconds), and it indicates that the accountant node of the current term can no longer provide adequate services, which means the accountant node may be offline due to network delay or system failure. At which time, the node that first discovers this situation becomes a candidate node and initiates an election for a new term accountant and sends a RequestVote RPC to all nodes in the network except itself. The nodes that receive a RequestVote RPC from the candidate node only respond to the vote request usually when the candidate node is not matched with RNL. The description of RequestVote RPC is shown in Table 9. When a candidate node gets half or more voting responses, the node changes from candidate state to accountant state and sends a heartbeat message to other nodes in the network. The purpose of this action is to prove that there is already an accountant node that can provide adequate services.

**Table 9. The description of RequestVote RPC in the RAC algorithm**

| | RequestVote RPC |
|---|---|
| **Parameter** | **Description** |
| Term | The current term of the candidate node. |
| CandidateID | The id of the candidate node. |
| **Return** | **Description** |
| Term | The current term of the node. |
| VoteGranted | Set to true when the candidate won this vote. |

④ $Package(Message, new\_Block, Accountant)$

The accountant node packages the requests from the client into a block with the structure shown in the Fig.1. A complete block message can be represented as < BlockNum, prehash, TimeStamp, Merkle-root, Entries[ ]>, together with the previous hash value, the timestamp, and other necessary messages, where Entries[ ] including the client transaction request. The block is then broadcast to the Evaluator Group to validate the block information. It is important to note that the original transaction request from the client is sent to the accountant node as well as to evaluator group at the same time.

⑤ $Verify(JudmentRPC, P_i, E)$

The evaluator node verifies the block information from the accountant node. The purpose of this step is to determine whether a Byzantine event has occurred. The Evaluator node first checks the block legitimacy information such as the hash value in the block. Secondly, it matches the Entries[ ] in the block to determine whether it is consistent with the transaction request from the client. Phase ⑤ is implemented through Judgment RPC between the node and the evaluator. The description of Judgment RPC is shown in Table 10.

**Table 10. The description of Judgment RPC in the RAC algorithm**

| | Judgment RPC |
|---|---|
| **Parameter** | **Description** |
| Term | The current term of the accountant node. |
| Accountant ID | The id of the accountant node |
| Entries[ ] | The transaction request that added by accountant node in the block. |
| **Return** | **Description** |
| Term | The current term of the node. |
| Success | Set to true when the transaction information from the accountant and that from client are equal. |
| Fail | Set to true when the transaction information from the accountant and that from client are not equal. |

⑥ $Re\_Blockchain(Accountant, AppendEntriesRPC, new\_Block)$

If the block is determined as invalid in phase ⑤, then the block information will be set to empty (More than 50% of the evaluator nodes respond to "Fail" to the accountant node in the Judgment RPC). The block information will be broadcast to all nodes after verification by the evaluator node. This process has interacted through AppendEntries RPC (The description of AppendEntries RPC is shown in Table 11). All nodes that receive the AppendEntries RPC will response "success" to the accountant node, indicating that the transaction request has been correctly added. When the block has been securely copied to more than 50% of the nodes, the account node will return the execution result to the client, thus completing the consensus of the request. The follower node will decide that the current accountant node is a byzantine node when they find an empty block added to the blockchain. Then the node will add the current leader node into their RNL and restart a new term of accountant node election.

**Table 11. The description of AppendEntries RPC in the RAC algorithm**

| | Judgment RPC |
|---|---|
| **Parameter** | **Description** |
| Term | The current term of the accountant node. |
| Accountant ID | The id of the accountant node |
| Entries[ ] | The transaction information that added by accountant node in the block. |
| **Return** | **Description** |
| Term | The current term of the node. |
| Success | Set to true when the transaction requests are successfully added to the local blockchain. |

We now give a full version of RAC to help better understand their decentralized design concepts, and the pseudocodes are presented in Algorithm 1.

```
Algorithm1 RAC
Input: Transaction Request from client.
Output: void

1. begin
2.   if no evaluator group in the permissioned blockchain system then:
3.       generate evaluator group
4.   end if
5.   while Transaction Request from client && evaluator group is existed do:
6.      if accountant node is not existed || the term of account node is less than current term || accountant node is byzantine node then:
7.          generate candidate
8.          communication between all Follower nodes and Evaluator group through RiskCompute RPC
9.          all node update RNL
10.         communication between candidate node and all follower nodes through RequestVote RPC
11.         if received vote from most follower nodes then:
12.             candidate node become accountant node
13.     else if accountant node existed then:
14.         add the Transaction Request into block
15.         communication between accountant node and Evaluator group through Judgment RPC
16.         if the number of "fail" in all Judgment RPC > 50% then:
17.             the block is set to empty block
18.         else then:
19.             the block is set to valid block
20.         communication between all Follower nodes and accountant node through AppendEntries RPC
21.         if follower node i find empty block then:
22.             add accountant node into RNL
23.             the term of follower node i is increase
24.   end while
25. end
```

## 4.2 Risk-Node Assessment Mechanism

It is necessary to evaluate and update the Byzantine nodes in the network to prevent Byzantine nodes from becoming accountant nodes. RAC describes the trustworthiness of a node by introducing the concepts of risk value and reliability. The risk value reflects the probability of Byzantine events occurring in a node during a specific term compared to other nodes in the network. Reliability reflects the probability of a Byzantine event occurring at the accountant node during their term. To improve the identification of malicious behavior, the Risk-Node Assessment Mechanism designed in this paper uses systemcall sequences as the source of data analysis. The system call represents the original interaction between the program and the host system, and there is no data abstraction in this process, the use of systemcall for intrusion detection system was first proposed by Forrest and assumes that program operation affects the system and that all exceptions leave traces in the system calls executed by the kernel [34]. Thus, the most significant advantage of our method is that it can find the occurrence of malicious behavior intuitively and locate malicious nodes with abnormal systemcall sequences quickly.

We offer some assumptions about the permissioned blockchain in order to set the stage for the discussion of the Risk-Node Assessment Mechanism.

Assumption 1: The number of honest nodes is always greater than that of Byzantine nodes in the network.

Assumption 2: The malicious behavior of Byzantine nodes cannot be predicted because attackers can use a variety of attack methods or tools to reach their goal of interfering with consensus individually or collusively.

Assumption 3: All nodes run on the same operating system, and the behavior of honest nodes is similar universally because they try to reach the consensus according to their responsibilities, which means that their systemcall sequence is also similar.

Based on the above assumptions, this section will use a short sequence-based technique to model the subsequence of system call sequences for honest nodes, and the node that deviates significantly from it will be considered malicious. The sequence of system calls for each node is first modeled. The sequence of system calls for all nodes is represented as a $num * k$ matrix, say $N_r$, where $num$ indicates the number of nodes in the network and each row in $N_r$ represents the entire sequence of system calls for a node in the previous term. Thus, the matrix $N_r$ is the matrix consisting of the short sequence of system calls of all nodes in the network. Combining the sequence of system calls with their frequency of occurrence is also necessary to highlight the significant differences in behavior between malicious nodes and honest nodes. We set $S_{(i,j)}$ to indicate the specific index of a sequence of system calls, $S_{(i,j)}$ is derived from Equation (6), and $f_s$ denotes the frequency of occurrence of $S_{(i,j)}$ in the matrix $N_r$.

$$S_{(i,j)} = T(i,j), \ 1 \leq i \leq num \ and \ 1 \leq j \leq k \tag{6}$$

$$f_s = \frac{S_{(i,j)}}{\sum_m S_{(m,j)}}, \ 1 \leq m \leq num \ and \ 1 \leq j \leq k \tag{7}$$

$$N_f(i,j) = S_{(i,j)} * f_s, \ 1 \leq i \leq num \ and \ 1 \leq j \leq k \tag{8}$$

Second, based on Assumption 3, normal nodes behave very similarly in the consensus process. To further reflect the deviation between the short sequence matrix composed of frequently used system call sequences and the short sequence matrix composed of less regularly used system call sequences, it is also necessary to pay attention to some system call sequences with low frequency but tremendous value because the critical information such as malicious attack is likely to be hidden in them. Therefore, the process of inverse document frequency is also required, and the detail is shown in Equation (9).

$$idf_{(i,j)} = log \frac{|num|}{|\{i:S_{(i,j)} \in N_i\}+1|}, \ 1 \leq i \leq num \ and \ 1 \leq j \leq k \tag{9}$$

In Equation (9), $|num|$ denotes the number of nodes, $|\{i:S_{(i,j)} \in N_i\}+1|$ is the number of nodes containing the specific system call sequence $S_{(i,j)}$. Finally, the original matrix $N_r$ is obtained as shown in Equation (10) after the processing of frequency and inverse document rate to obtain the matrix $N_f$.

$$N_f(i,j) = S_{(i,j)} * f_s(i,j) * idf_{(i,j)}, \ 1 \leq i \leq num \ and \ 1 \leq j \leq k. \tag{10}$$

Assumption 1 and Assumption 2 can analyze the risk value using an unsupervised outlier detection algorithm. In this paper, we use the isolated forest algorithm [35] for malicious node determination. The purpose of the isolation forest algorithm is to rank each node by its anomaly score. The isolation forest consists of many binary trees, each of which is called an isolation tree. The construction of an isolation tree is an entirely random process. Let there be $num$ nodes in the network, the method of estimating the average path length of the leaf nodes of an isolation tree can be referred to as the unsuccessful search of a binary tree because they have a very similar structure, as shown in Equation (11).

$$c(n) = 2H(n-1) - 2\left(\frac{n-1}{n}\right) \tag{11}$$

Where H(x) is the Euler–Mascheroni constant, $H(x) = \ln(x) + 0.5772156649)$. It can be used to normalize the path length because $c(n)$ is the average of the path length. The anomaly score of a node $x$ can be calculated by equation (12), where $h(x)$ is the path length of node $x$ in the isolated tree.

$$s(x,n) = 2^{-\frac{E(h(x))}{c(n)}} \tag{12}$$

The risk-node assessment mechanism can calculate the risk values of each node by analyzing the sequence of system calls of the nodes in the previous term. The node will be judged as a Byzantine node when its risk value is significantly greater than other honest nodes.

**4.3 Punishments Mechanism**

This section will focus on the penalty mechanisms in the consensus algorithm. The symbol glossary is listed in the Table 12 to facilitate expressing the penalty mechanisms. We use DFA (Deterministic Finite Automaton) to describe the Byzantine node. A DFA is a quintuple $< S, \Sigma, \delta, S_0, F >$, where $S$ is a finite set of states, $S_0$ is the initial state, $F$ is a set of acceptable states, $\Sigma$ is a finite set of alphabets, $\delta$ is conversion function, $\delta$ can be expressed as $\delta = s \times \Sigma \to S$.

Table 12. Comparisons between two types of consensus algorithm in blockchain

| Symbol | Description |
| --- | --- |
| Action | The set of operations performed by the follower, accountant, evaluator in different phase of the RAC. Action = {receive, generate new block, broadcast, verify, send system call ...}. |
| Trace | A sequence on the set Action. Such as {receive → generateNewBlock → broadcast → valid block}. |
| Behavior | The behavior of node 1 can be denoted as <1, tr>, the identification of the node is 1, and tr is a Trace. |

As shown in Fig.6 (a), we define the Distinguishing Automaton for the behavior of an accountant node, in which the initial state is 0 (S = 0), acceptable states are [4, 5] (F ∈ {4,5}), $\Sigma$ is the action, where "receive" indicates the receive the transaction request from the client, "generate new block" is the action of the convert transaction request to new block, "broadcast" means broadcast the new block to the evaluator group according to the description of phase ③, "valid block" and "empty block" represents the result of the block generated after verification of evaluator group ($\Sigma \in$ {receive, generate new block, broadcast, valid block, empty block}).

As shown in Fig.6 (b), we define the Distinguishing Automaton for the behavior of an evaluator node, in which, the initial state is 0 (S = 0), acceptable states are [3, 4] (F ∈ {3,4}), $\Sigma$ is the action ($\Sigma \in$ {receive, verify, success, fail}), where "receive" indicates receive the transaction request from the client and the block from the accountant node in phase ⑤, "verify" means verify the block from the accountant node, "success" is thoughted that the evaluator has made a right decision, "fail" represents the decision of a new block by an evaluator is wrong because their judgment is the opposite of that of the majority.

As shown in Fig.6 (c), we define the Distinguishing Automaton for the behavior of a follower node, in which, the initial state is 0 (S = 0), acceptable states are [4, 5] (F ∈ {4,5}), $\Sigma$ is the action, where "receive" indicates receive the block that from the accountant node and verified by the evaluator group, "addition new block" means that add this block to the local blockchain of the follower node in phase ⑥ and the behavior of honest nodes is similar universally in this phase, "send systemcall" is the action that sends systemcall of previous to the evaluator group in phase ①, "abnormal" and "normal" represents the result of risk-node assessment mechanism executed by the evaluator group in phase ①. Where Σ ∈ {receive, addition new block, send systemcall, abnormal, normal}.

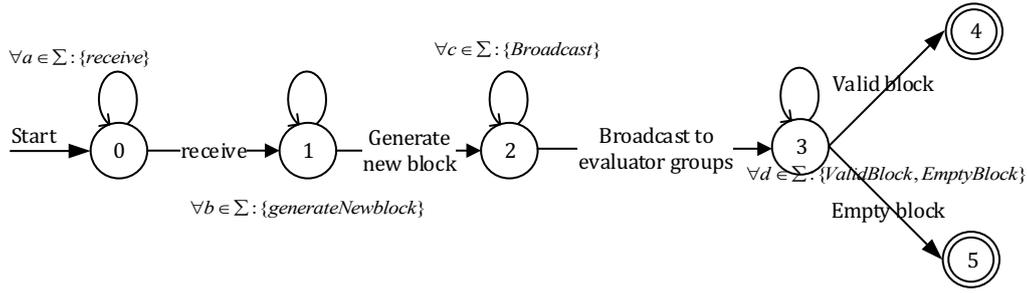

(a). the state graph of accountant node

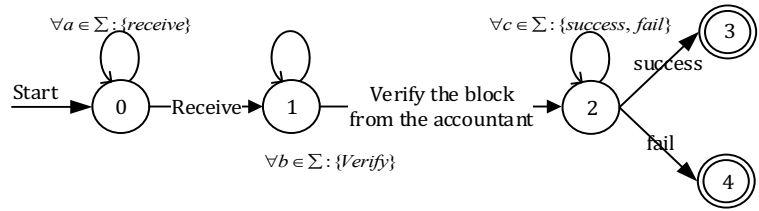

(b). the state graph of evaluator node

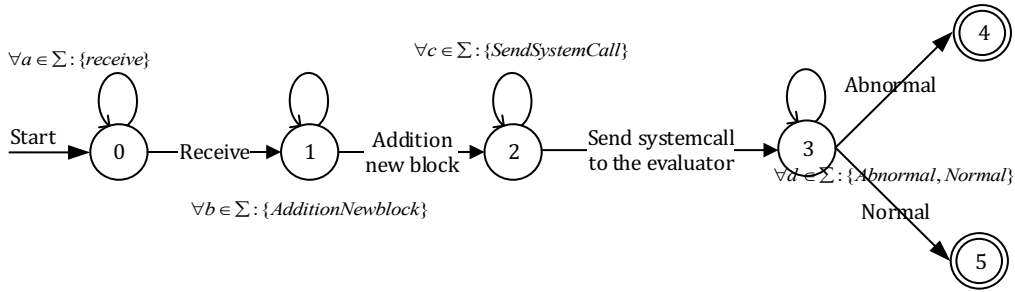

(c). the state graph of follower node

**Fig.6. The state graph of RAC**

Criterion 1: Byzantine Behavior of Accountant. Let <1, tr> denote the behavior of accountant node 1. It will be considered that node 1 is a byzantine node when node 1 reaches state 5 because the new block generated by the accountant node is judged to be invalid in phase ⑤, which may indicate that the transaction request from the client was maliciously tampered with by accountant.

Criterion 2: Byzantine Behavior of Evaluator. Let <1, tr> denote the behavior of evaluator node 1. It will be considered that node 1 is a byzantine node when node 1 reaches state 4 because their judgment is the opposite of that of the majority in phase ⑤, which may indicate that collusion attacks happened in the accountant node and the evaluator node.

Criterion 3: Byzantine Behavior of Follower. Let <1, tr> denote the behavior of follower node 1. It will be considered that node 1 is a byzantine node when node 1 reaches state 4 because their behavior of block addition is different from that of the majority in phase ⑥, which will be decided in the risk-node assessment mechanism.

Node i was judged as a byzantine node according to Criterion 1, Criterion 2, and Criterion 3. The byzantine node will be added in RNL to prevent it from causing more significant losses to the blockchain system, which means that it can only serve as the follower node. It is an acceptable solution that the organization of the Byzantine nodes needs to pay a portion of stake interests or fines to reward those who had honest nodes. In addition, some compensation mechanisms can be added based on agreements reached between organizations in permissioned blockchain. For example, the organizations of the honest nodes can remove the byzantine node from RNL through the smart contract after receiving the reward.

## 5. PERFORMANCE AND EVALUATIONS
### 5.1 Theoretical Analysis
#### 5.1.1 Complexity of consensus

Every two nodes' communication between each other and a three-stage commit process was required in the PBFT algorithm before adding blocks to the blockchain, which leads to a high communication complexity of PBFT. However, the state classification mechanism of the RAC algorithm makes it possible to add blocks without high communication costs.

We divide the RAC algorithm into three main phases to facilitate the analysis: accountant selection stage (including phase ①, phase ② and phase ③ of RAC algorithm), block addition stage (including phase ④ and phase ⑤ of RAC algorithm) and transaction confirmation stage (including phase ⑥ of RAC algorithm). Let $n$ denote the number of nodes, $n_e$ indicates the number of evaluator nodes, $n_f$ is the number of follower nodes.

In the accountant selection stage, the follower node needs to go through two steps of sending systemcall to the evaluator group and sending voting request to the rest of the follower nodes when it becomes an accountant node, so the communication complexity of this stage is $O(n + n_f)$. In the block addition stage, only the accountant node needs to send messages to the evaluator node and the follower node to reach consensus. No communication is needed between the evaluator nodes and follower nodes, so the communication complexity of the RAC algorithm at this stage is $O(n_e + n_f)$. The communication complexity of RAC algorithm in the transaction confirmation stage is $O(1/2\, n)$ because only more than half of the follower nodes are required to complete the verification. Therefore, the communication complexity of the RAC algorithm is $O(2n_f + 3/2\, n + n_e)$, which is smaller than $O(n^2)$ in PBFT.

#### 5.1.2 Security analysis

Some malicious attacks can threaten the permissioned blockchain. In this section, we situate our discussion about the security ability of the RAC algorithm.

Collusion attack and Double Spending attack: It would have an unpredictable impact on the blockchain system if the consensus protocol were not protected against double-spending attacks. The consensus result of the RAC algorithm is deterministic rather than probabilistic, which can avoid double-spending attacks to a certain extent. However, double-spending attack occurs when the accountant and the malicious evaluator collude in the RAC algorithm. Thus, the defense mechanism for the double-spending attack can also be equivalent to that for the collusion attack. Collusion and double-spending attacks cannot occur if the percentage of malicious evaluator nodes does not exceed 50% in the RAC algorithm because blocks can only be added after more than 50% of the evaluator nodes have passed the checksum, which also means that the RAC algorithm can tolerate at most 49% of Byzantine evaluator nodes in the network. In the actual blockchain network, the attacker may use the 51% attack as a sub-attack of the double-spend attack or collusion attack.

Sybil attack: Sybil attack indicates that the attacker attempts to control the network by creating many fraudulent identities, which can be used to help the attacker gain voting power or broadcast false messages in the blockchain system. No node of the decentralized system can know how many nodes are involved in the peer network; they can only judge the global situation by the data they receive. The attackers can take the means to maliciously control the honest nodes by disguising themselves as multiple nodes and influencing the behavior of the honest nodes. Two approaches are used in the RAC algorithm to defend against Sybil attacks. On the one hand, nodes need to complete registration in the identity authentication mechanism and be assigned a unique identity before joining the permissioned blockchain. Registration needs to be endorsed by a specific organization, which can enhance the cost for attackers to register multiple identities in the network maliciously. On the other hand, the RAC algorithm uses Risk-Node Assessment Mechanism to defend against Sybil attack when identity authentication mechanism was bypassed by the attacker, the node where the abnormal behavior of forging identity occurs will be readily determined by the Risk-Node Assessment Mechanism because there will be a massive difference between the byzantine node and the honest node in their systemcall record.

Targeted Attack: Some consensus algorithms are vulnerable to Targeted Attacks because the accountant node can be inferred. For example, an attacker launches a DOS attack on the current accountant node, causing it to be inoperative until a malicious node controlled by the attacker becomes the accountant. Thus, the method that periodic or irregular accountant replacement is used to defend against Targeted Attack, this approach is only valid under the assumption that the attacker cannot damage the accountant node immediately. RAC algorithm can achieve defense against Targeted attacks in an effective way due to the function of term.

The accountant node cannot be directly generated from follower nodes but needs to go through the candidate state and be voted by more than half of the nodes before becoming an accountant node. A term is divided into accountant election and normal operation, the only one accountant node existed in the certain term, term can keep the real identity of accountant uncertain in the accountant election stage until the accountant providing services for blockchain networks in normal operation stage. But when the accountant node is elected successfully, it is too late to attack the accountant in this term because the accountant node has been able to serve normally without malicious controlled by attacker. If attacker wants to perform a targeted attack on the permissioned blockchain, the only thing it can do is to corrupt on the node randomly before the next term. The randomness of different network environments and latency of each node makes it impossible for the attacker to predict which node will become the accountant node in the RAC algorithm. Therefore, it's totally random behavior with the probability that can be calculated in the Equation (13).

$$P_1(Attack\ the\ accountant\ of\ next\ term) = 1/N \qquad (13)$$

In addition, an attacker who wants to take malicious control of a node will generate some traces of attack chain in this node, such as Reconnaissance (Use social engineering to understand the target), Weaponization (Targeted attack tool creation), Delivery (Delivery of the attack tool to the target node), Exploitation (The attack tool is triggered on the node to realize controlled based on the system's application or operating system vulnerabilities), Installation (Remote control program of the installation) and Command & Control (The C2 channel be established between the compromised node and internet controller server).

Therefore, successful malicious control of a node by an attacker will cause the systemcall of that node to be completely different from other nodes in the permissioned blockchain. Node Assessment Mechanism in the RAC algorithm can detect malicious behavior because the systemcall sequence of all nodes needs to be analyzed by the evaluator group before accountant node election in the new term. The probability $P_2$ can be calculated by equation (14), where $Pr$ indicates the number of nodes that be malicious controlled to becoming accountant node by attackers

$$P_2 = \lim_{pr \to 0} Pr/N \tag{14}$$

### 5.1.3 Summary of theoretical analysis

We assume that there are $n$ nodes in the network and $x_{i,j}$ indicates the $j$-th transaction in the $i$-th block. $n_d$ is the number of failed nodes due to downtime or network errors. The function $f(x_{i,j}) \to \{0,1\}$ is used to denote the result of consensus, where 0 presenting invalid and 1 presenting valid of each transaction to reach a consensus. We can draw the following conclusions about the theoretical performance of the RAC algorithm and summarize them in Table 13 based on the previous analysis.

Agreement. Consensus can be reached normally when the number of failed nodes is less than 50% of the total number of nodes in the RAC algorithm. $f(x_{i,j}) = 1$, when $n_d/n < 50\%$.

Efficiency. The communication complexity required to reach consensus in the RAC algorithm is $O(n)$.

Security. The RAC algorithm can form an effective defense against Sybil attack, Targeted Attack, Collusion attack and Double Spending attack when the number of malicious nodes is satisfied to be smaller than 50%.

Table 13. Comparisons between two types of consensus algorithm in blockchain

| Consensus algorithms | Complexity of consensus | Crash Fault Tolerance | Byzantine Fault Tolerance | Sybil attack | Targeted Attack |
|---|---|---|---|---|---|
| PBFT | $O(n^2)$ | 33% | 33% | Vulnerable | Vulnerable |
| Tendermint BFT | $O(n^2)$ | 33% | 33% | Vulnerable | Vulnerable |
| RAFT | $O(n)$ | 33% | / | Vulnerable | Vulnerable |
| CRAFT | $O(n)$ | 51% | 51% | / | Vulnerable |
| Beh-Raft | $O(n)$ | 51% | 51% | Safe | / |
| RAC | $O(n)$ | 51% | 51% | Safe | Safe |

### 5.1.4 Comparison with the improved consensus algorithms

This section compares RAC with some improved consensus algorithms of permissioned blockchains, such as Beh-Raft, HHRAFT, CRAFT, and Tendermint BFT, in terms of decentralization security and scalability under the evaluation system introduced in section 3.1 to highlight the advantages of RAC.

We set there are $n$ nodes in the permissioned block network. The preliminary evaluation results are shown in Table 14. An explanation of each indicator can be found in section 2.1.

Table 14. Preliminary Evaluation Results

| Indicator \ Algorithm | Beh-Raft | HHRAFT | CRAFT | Tendermint BFT | RAC |
|---|---|---|---|---|---|
| Number of consensus nodes | Part | Part | All | All | Part |
| Accountant selection method | Voting | Voting | Voting | Polling | Voting |
| Consensus nodes weight | No | Yes | No | Yes | Yes |
| Byzantine fault tolerance | 51% | 16% | 51% | 33% | 51% |
| Byzantine node controllability | Yes | Yes | No | No | Yes |
| Attack behavior costs | Middle | Low | Low | High | Middle |
| Resource consumption | $O(n)$ | $O(n)$ | $O(n)$ | $O(n^2)$ | $O(n)$ |

Based on the analysis results in Table 10, combined with the quantitative evaluation criteria introduced in section 3.1, evaluation matrix $\boldsymbol{B} = (b_{ij})_{5\times 7}$ is constructed as follows.

$$\boldsymbol{B} = \begin{bmatrix} 0 & 0 & 1 & 1 & 0 \\ 0.5 & 0.5 & 0.5 & 1 & 0.5 \\ 0 & 1 & 0 & 1 & 1 \\ 0.7 & 0.3 & 0.7 & 0.5 & 0.7 \\ 1 & 1 & 0 & 0 & 1 \\ 0.6 & 0.3 & 0.3 & 1 & 0.6 \\ 0.7 & 0.7 & 0.7 & 0.3 & 0.7 \end{bmatrix}$$

Positive ideal solution is $C^+ = [1, 1, 1, 0.7, 1, 1, 0.7]$ and Negative ideal solution is $C^- = [0, 0.5, 0, 0.3, 0, 0.3, 0.3]$. We compared the proximity between ideal solution of Beh-Raft, HHRAFT, CRAFT, Tendermint BFT and RAC. The experimental results are shown in Fig.7.

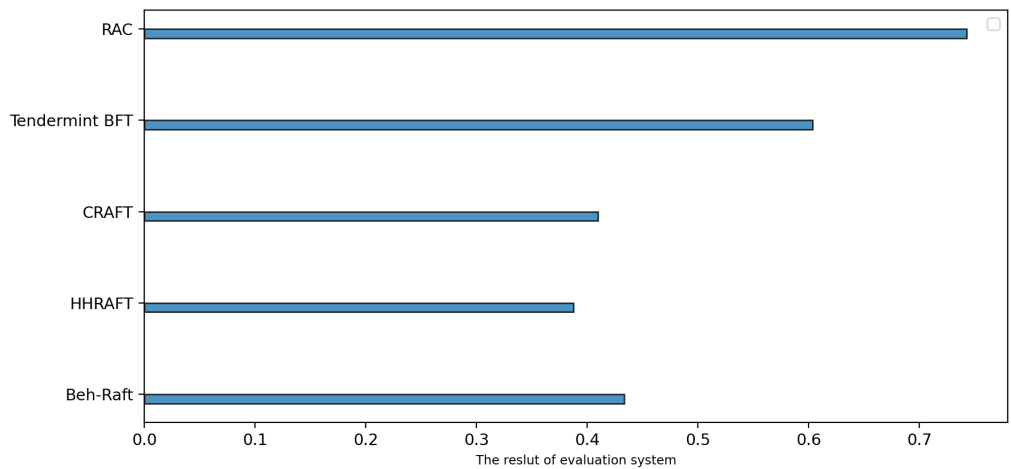

**Fig.7. The experimental results of Comparison with the improved consensus algorithms**

Under the mathematical model of blockchain evaluation proposed in section 3.1, the RAC algorithm has a more significant advantage over other improved consensus algorithms.

**5.2 Experiment Analysis**

The performance of RAC algorithms can be experimentally compared and analyzed. The RAC algorithm combines Byzantine fault tolerance and high transaction efficiency, making it applicable to real-world network environments. We conducted the experiments using an Intel Core i5 and 16 GB RAM running macOS operating system. The construction of the RAC algorithm is implemented using Golang1.14.7. To further test the performance of the proposed approach in a cluster environment, we use the technology of container, thread, and virtual machine to represent the different network nodes, the technology of container, thread, and virtual machine are implemented with goreman0.3, docker18.09, and VMware fusion11.5.5. We compare the RAC with consensus algorithms such as Tendermint BFT and Raft that are widely used in blockchain platforms available today in efficiency performance.

### 5.2.1 The comparison of election cost

Accountant selection is the first phase in the consensus algorithm. The accountant selection of RAC and Raft is based on voting. Fig.8. depicts the comparison between Raft and RAC algorithms for different network sizes in the accountant election cost. The RAC algorithm consumes more cost of accountant election than the Raft algorithm due to the risk-node assessment mechanism. The cost of accountant selection has experimented in a network environment with 10, 20, 30 nodes and different percentages of Byzantine nodes. The results show that as the network sizes and the proportion of Byzantine nodes increase, the cost of the RAC increase more than that of the Raft algorithm.

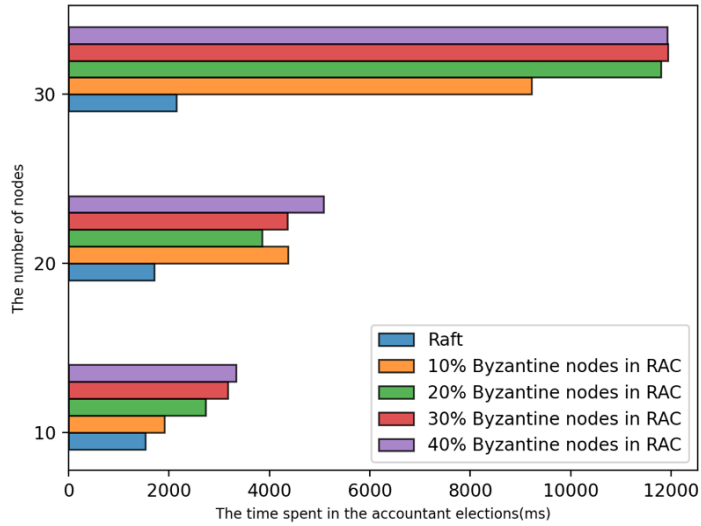

Fig.8. The cost of accountant election over different network size

### 5.2.2 The comparison of transaction confirmation latency

Transaction confirmation latency is when a specific transaction reaches consensus in all nodes, which is one of the essential indicators of a consensus algorithm. It is calculated as shown in Equation (15).

$$Latency = time_{block} - time_{request} \qquad (15)$$

Where $time_{block}$ denotes the timestamp when a transaction reach consensus, and $time_{request}$ indicates the timestamp when the transaction is initiated. The few slow nodes in both RAC and Raft algorithms do not affect the overall latency performance of the algorithm. However, compared to the Raft algorithm, the RAC algorithm needs to add a message forwarding process. Due to their unique judgment mechanism of the new block, some latency performance was affected. Fig.9. depicts the latency performance of RAC and Raft algorithms with the different number of nodes. The overall latency performance of the RAC algorithm is reduced by about 50% in a network environment with 5-30 nodes.

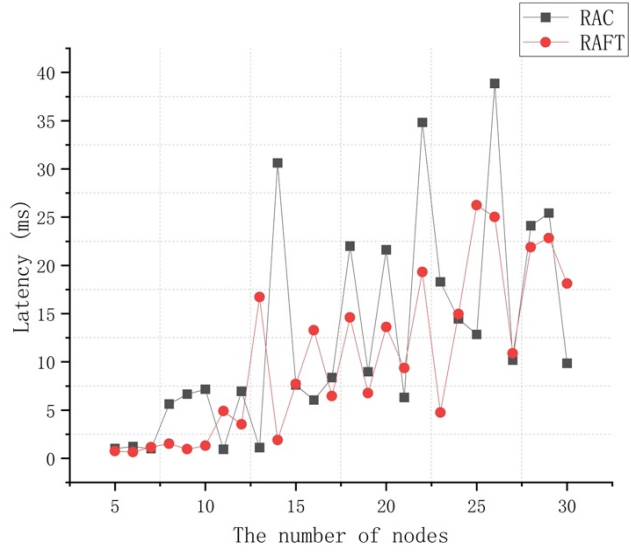

**Fig.9. Comparison of latency between Raft and RAC for 5-30 nodes**

**5.2.3 The comparison of transaction throughput**

Throughput indicates the ability to process the number of client requests per unit time, and Equation (16) is the formula for calculating throughput, which can measure the efficiency of the blockchain system.

$$Throughput = \frac{number\ of\ transactions\ T}{the\ time\ of\ T\ transactions\ reaching\ a\ consensus} \quad (16)$$

The value of $T$ is taken as 1000, and the experimental results of throughput performance comparison of different consensus algorithms are shown in Fig.10. In a network environment with 5-30 nodes, the results show that the RAC algorithm loses about 10% of throughput performance due to its security mechanism when compared to the Raft algorithm. However, compared with the Byzantine fault-tolerant consensus algorithm such as PBFT, the RAC algorithm still has an advantage in terms of throughput performance.

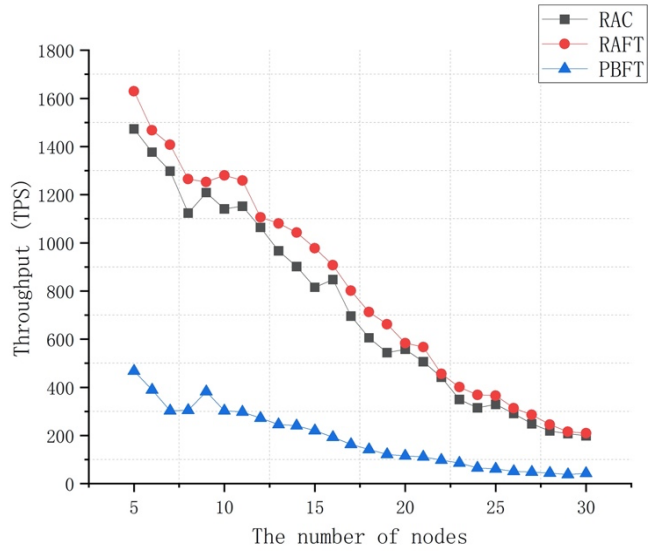

Fig.10. Comparison of throughput between Raft, PBFT and RAC for 5-30 nodes

### 5.2.4 The comparison of time efficiency

We compare the time efficiency of different consensus algorithms under the simulation environment that uses a 100-node to generate concurrency transactions, the size of every block generated is 256KB. As shown in Fig.11. the time efficiency of RAC at a better level than Tendermint BFT. Under the same size of blocks, RAC are superior to Tendermint BFT due to communicational complexity of Tendermint BFT is up to $O(n^2)$, while RAC have reduced it to a smaller value, as a result, the time consumed of Tendermint BFT is more than RAC with the large number of transactions.

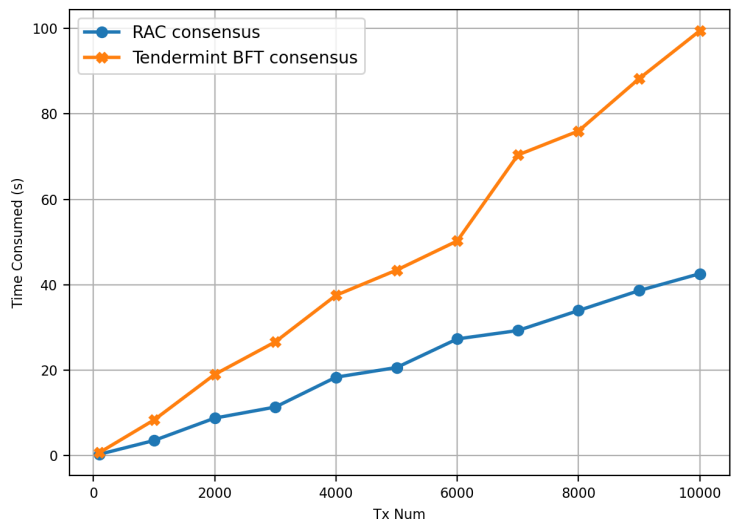

Fig.11. Comparison of time efficiency between Raft and Tendermint BFT

### 5.2.5 Summary of experiment analysis

Compared to the Raft algorithm, the RAC requires an additional security mechanism, so there is a loss of efficiency in the simulation environment. However, RAC has advantages in latency and throughput performance compared to the

Byzantine fault-tolerant consensus algorithm like Tendermint BFT. Thus, our results show that the RAC algorithm proposed in this paper can reach a better performance balance in efficiency and security.

## 6. CONCLUSIONS AND FUTURE WORKS
### 6.1 Conclusions

We propose a consensus algorithm, RAC, for permissioned blockchain in this paper. First, we introduce the implementation process of the RAC algorithm, which focuses on the detailed phases, Risk-Node Assessment Mechanism, Reward and Punishments Mechanism of the algorithm. Second, we theoretically analyze the performance of the RAC algorithm and compare the RAC algorithm with other improvement consensus algorithms. The result shows that the RAC algorithm has high scalability (the communication complexity of the RAC algorithm is $O(n)$) and can provide better protection against blockchain attacks such as Sybil attacks and Targeted attacks. Finally, we experimentally compare the efficiency performance of the RAC algorithm with the consensus algorithms (Raft and Tendermint BFT) that are widely used in real-world blockchain systems. The results show that the RAC algorithm can achieve lower latency performance and higher throughput under a Byzantine network environment (the RAC algorithm loses about 10% of throughput performance compared to the Raft algorithm; however, Raft can be only used in a non-byzantine environment, RAC algorithm still has an advantage in terms of throughput performance compared with Tendermint BFT under Byzantine network environment). In summary, RAC can reach a better balance in performance of decentralization, security and scalability.

### 6.2 Future work

From what has been concluded in theoretical and experimental analysis, RAC can be well applied to the new business model based on blockchain to help a group of entities under cooperation and wants to get rid of the dependence on centralization certification organization. Table 15 lists some real-world applications of permissioned blockchain where RAC can be selected as the consensus algorithm. These scenarios are characterized by fixed node size and the presence of a certain number of light nodes but with certain security and time efficiency performance requirements. In the future, we will further use RAC in some complex scenarios and explore the performance of RAC on different real-world applications.

Table 15. The application scenarios of RAC can be used

| Scenario | Participating methods | Characteristics | Node size |
|---|---|---|---|
| Supply chain | Internal cooperation of some companies | Medium time efficiency<br>Many light nodes<br>Node size fixed | <100 |
| Smart cities | IoT application | Medium time efficiency<br>Many light nodes<br>Node size fixed | <1000 |
| Vehicular ad-hoc networks | IoT application | Medium time efficiency<br>Many light nodes<br>Node size fixed | <1000 |
| Smart home | IoT application | Medium time efficiency<br>Some light nodes<br>Node size fixed | <20 |
| Healthcare system | Enterprises cooperate to provide external services | Medium time efficiency<br>Some light nodes<br>Node size fixed | <100 |

Even though RAC is an improved consensus solution in permissioned blockchain, it is still some limitations:

(1) RAC has a lower processing speed than traditional centralized systems, which may limit the application of the RAC algorithm.

(2) Reputation-based incentive mechanism should be established and enforced to reduce the technical barriers and costs of the punishments mechanism we proposed in RAC.

(3) The security of RAC needs to improve by developing some new technologies. The Risk-Node Assessment Mechanism of RAC introduces the threat of re-centralization, which destroys the distributed security features of the blockchain system to a certain extent, which is conducive to the implementation of 51% attacks and double-spending attacks.